\documentclass[showpacs,twocolumn,pre,superscriptaddress]{revtex4}

\def\n{\langle n \rangle}
\def\lappeq{\mathrel{\rlap{\raise.5ex\hbox{$<$}}
{\lower.5ex\hbox{$\sim$}}}}

\usepackage{graphicx}
\begin{document}
\title{Nonequilibrium wetting transitions with short range forces}
\author{F. de los Santos} 
\affiliation{Center for Polymer Studies and Department of Physics, 
Boston University, Boston, MA 02215 USA}
\affiliation{Departamento de F{\'\i}sica da Faculdade de Ci{\^e}ncias e  
Centro de F{\'\i}sica da Mat{\'e}ria Condensada da Universidade de  
Lisboa,\\ Avenida Professor Gama Pinto, 2, P-1643-003 Lisboa Codex, 
Portugal}
\author{M.M. Telo da Gama} 
\affiliation{Departamento de F{\'\i}sica da Faculdade de Ci{\^e}ncias e  
Centro de F{\'\i}sica da Mat{\'e}ria Condensada da Universidade de  
Lisboa,\\ Avenida Professor Gama Pinto, 2, P-1643-003 Lisboa Codex, 
Portugal}
\author{M.A. Mu\~noz}
\affiliation{Departamento de Electromagnetismo y F\'\i sica de la Materia,
Universidad de Granada, 18071 Granada, Spain}

\begin{abstract}
We analyze within mean-field theory as well as
numerically a KPZ equation 
that describes nonequilibrium wetting. 
Both complete and critical wettitng transitions were found and
characterized in detail. For one-dimensional substrates the critical
wetting temperature is depressed by fluctuations.  
In addition, we have investigated a region in the space of parameters 
(temperature and chemical potential) where the wet and 
nonwet phases coexist.  
Finite-size scaling analysis of 
the interfacial detaching times indicates that the
finite coexistence region  
survives in the thermodynamic limit. 
Within this region we have observed
(stable or very long-lived) structures related to 
spatio-temporal intermittency in other systems. In the interfacial  
representation
these structures exhibit perfect triangular (pyramidal) patterns 
in one (two dimensions), that are characterized by their slope and size 
distribution.

\end{abstract}


\maketitle
\section{Introduction}

When a bulk phase $\alpha$ is placed into contact with a  
substrate, a layer of a second, coexisting, phase $\beta$ may form 
if the substrate preferentially adsorbs it. At a {\it wetting transition},
the thickness of the $\beta$ layer diverges.
Equilibrium wetting has been experimentally observed and theoretically 
investigated using, among many other techniques, 
interface displacement models \cite{margarida,fisher,bhl}. 
Within this approach one considers the local height of the 
$\alpha \beta$ interface measured from the substrate, $h({\bf x})$, and
constructs an effective interface Hamiltonian, ${\cal H}(h)$ 
\cite{nota1}.
In equilibrium situations, one typically has
\begin{equation} {\cal H}(h)=\int_0^\infty d{\bf x}\Bigg[ 
{1 \over 2} \nu (\nabla h)^2 +V(h)\Bigg].
\label{ham}
\end{equation}
where $\nu$ is the interfacial tension of the $\alpha \beta$
interface (or the interfacial stiffness if the medium 
is anisotropic) and $V(h)$ accounts for the interaction
between the substrate and the $\alpha \beta$ interface. 

If all the microscopic interactions are short-ranged, one may take 
for sufficiently large $h$ at bulk coexistence \cite{bhl} 
\begin{equation}
V(h)=b(T)e^{-h}+ce^{-2h},
\label{potential} 
\end{equation}
where $T$ is the temperature, $b(T)$ vanishes linearly  
as $T-T_W$, $T_W$ being the wetting temperature,
and $c>0$ \cite{nota2}. By minimizing (\ref{ham})
one finds \cite{margarida,bhl} a {\em critical wetting} 
transition at $b=0$, i.e.
the interface height (or equivalently, the
wetting layer thickness), $\langle h \rangle$, diverges continuously as 
$b \sim T-T_W \to b_W=0^-$.
Equilibrium critical wetting has been
studied for decades and a rich (non-classical) behavior predicted 
\cite{margarida,fisher}. 

Wetting transitions may also be driven by the chemical potential
difference between the $\beta$ and $\alpha$ phases, $\mu$.
In this case wetting occurs 
at any temperature above $T_W$ (i.e. for $b>b_W$)
as $\mu =0$ is approached from the $\alpha$ phase. 
This is always a continuous transition and it is known as 
{\em complete wetting} \cite{margarida}.
A study of complete wetting requires adding a linear
term $\mu h$ to the Hamiltonian (\ref{ham}).  

A dynamic model for the growth of wetting layers 
has been proposed through the Langevin equation \cite{lipowsky}

\begin{equation}
\partial_t h({\bf x},t) = -{\delta {\cal H} \over \delta h}+\eta=
\nu \nabla^2h -{\partial V \over \partial h} + \eta,
\label{eweq}
\end{equation}
where $\eta$ is Gaussian white noise with mean and
variance 
\begin{eqnarray}
\langle \eta ({\bf x},t)\rangle &=&0, \nonumber \\
\langle \eta({\bf x},t) \eta({\bf x}',t') \rangle &=& 2 D \delta(t-t') 
\delta ({\bf x}-{\bf x}').
\label{noise}
\end{eqnarray}
Equation (\ref{eweq}) is an Edwards-Wilkinson (EW) growth equation
\cite{ew,barabasi} in the presence of an effective interface potential
$V(h)$. It describes the relaxation of the interfacial height $h$ towards
its equilibrium value, i.e. the value of $h$ that minimizes $\cal H$.
Within this context $\mu$ can be viewed as an external driving force
acting on the interface. Recall that in the absence of the wall, 
the corresponding equilibrium states for $\mu<0$
and $\mu>0$ are the $\alpha$ and $\beta$ phases, respectively, whereas 
phase coexistence occurs at $\mu=0$. 

Equilibrium models, however, are not sufficient to study wetting in
Nature,
since in a wide range of phenomena ({\em e.g.\/}, growth of
thin films, spreading of liquids on solids or crystal growth) thermal
equilibrium may not hold.
Nonequilibrium wetting transitions have been recently
studied by Hinrichsen et al. \cite{haye1}
in a lattice (restricted solid-on-solid) 
model with dynamics that do not obey detailed balance. 
The continuum nonequilibrium counterpart of this discrete model,
is a Kardar-Parisi-Zhang (KPZ) equation in the presence of a 
bounding potential, whose properties have been analyzed 
extensively by one of the authors and collaborators \cite{MN,genovese}. 
Clearly this is the most natural extension of
the EW equilibrium growth model to non-equilibrium situations.
In fact, in the absence of a substrate the KPZ non-linearity, 
$\lambda (\nabla h)^2$, is generically the most relevant nonequilibrium 
perturbation to the equilibrium EW equation \cite{barabasi}.
The KPZ nonlinearity is related to lateral growth, and 
although this mechanism is unlikely to be relevant in 
simple fluids, it may determine the wetting behavior of systems
with anisotropic interactions for which
the growth of tilted interfaces depends on their orientation \cite{wiese}.
For instance, it has been shown that crystal growth 
from atomic beams is described by the KPZ equation \cite{villain}. 
From a theoretical point of view a key and ambitious
task is that of developing general criteria to establish whether the KPZ
non-linear term should be included in a given interfacial model.

Related works published recently by
M\"uller et al. \cite{muller}, Giada and Marsili 
\cite{marsili}, Hinrichsen et al. \cite{haye2},
and ourselves \cite{nos}, consider similar non-equilibrium models 
in the  presence of various types of walls. 

In this paper, we further study the KPZ interfacial equation
in the presence of different types of potentials, attractive 
and repulsive. We 
focus on the connection of the associated phenomenology with 
non-equilibrium wetting and depinning transitions. 
In particular, we will stress that the transitions called
``first-order non-equilibrium wetting'' in \cite{haye2} are
not wetting transitions but, rather, depinning transitions. 
Also, we study for the first time the two-dimensional version
of this model, and report on new phenomenology.

The remainder of this paper is organized as follows. In the 
next section we introduce our KPZ-like model. A mean-field
picture is provided in section III, and its predictions 
numerically tested in section IV. The
conclusions are summarized in the final section.

\section{The model}

The model under study is defined by the Langevin equation
\begin{equation}
\partial_t h({\bf x},t)= \nu \nabla^2 h +\lambda (\nabla h)^2
-{\partial V(h)\over \partial h} + \eta, 
\label{kpzyv}
\end{equation}
where $V(h)= -(a+1)h + b  e^{-h}+c e^{-2h}/2$, $c>0$,
$\eta$ obeys (\ref{noise}) and 
$ a+1 =\mu $ is a chemical potential.
  
In the absence of the exponential terms (the limiting  wall) 
the interface moves with a nonzero positive (negative) mean velocity
 for $\mu$ larger (smaller) than a certain critical value $\mu_c$. 
 In one dimension $\mu_c$ can be found analytically since 
both the KPZ and EW equations have the same Gaussian steady-state 
height distribution \cite{barabasi}. 
Thus $\mu_c= \lambda \langle(\nabla h)^2\rangle$, which for discrete lattices 
can be approximated by \cite{krug} $\mu_c=-(D \lambda) /(2\nu \Lambda)$,
 $\Lambda$ being the lattice cutoff. 
Note that for $\lambda \not=0$ a nonzero chemical potential 
is required to balance the force exerted by the nonlinear 
term on the tilted interfacial regions. 
For $\lambda =0$ the model reduces to the equilibrium 
one and $\mu_c=0$ as usual. 
For negative values of $\lambda$ the interface is (on average)
pushed against the wall, while for positive $\lambda$ it is 
pulled away from the wall.
Thus the behavior of the system is determined by the sign of $\lambda$
\cite{note}.
In this paper we will consider $\lambda <0$ only (which corresponds to
the case studied using microscopic models \cite{haye2}).
Results for positive values of $\lambda$ will be 
published elsewhere. 

It is our purpose to study the effects of a substrate 
that adsorbs preferentially one of the two phases
on the stationary properties of the interface. 
This is achieved by considering
$b$ to be negative in equation (\ref{kpzyv}). 

This same equation, (\ref{kpzyv}), has been recently studied by Giada {\em
et al.\/} \cite{marsili} as a generic nonequilibrium continuum model for
interfacial growth.  
However, their choice of control parameters \cite{marsili}
privileges the role of the noise as the driving force of 
the nonequilibrium transitions. By contrast, motivated by the role
of the chemical potential and temperature in equilibrium
wetting, we fix the noise intensity and choose $a$ and $b$ 
as control parameters that are the fields driving critical and
complete wetting transitions.

To establish the analogy  with equilibrium, let us stress 
that just like in {\it equilibrium complete wetting},
{\it nonequilibrium complete wetting} occurs when the attractive 
potential $V(h)$ is not capable of binding the interface, 
at temperatures above the wetting
transition temperature, $b > b_W$, as the chemical potential approaches
that of `bulk coexistence', $\mu \to \mu_c$.
At this transition the interface begins to move and $\langle h \rangle$
diverges. On the other hand, the {\it nonequilibrium} analogue of
{\it critical wetting} corresponds to the unbinding of a bound interface at
`bulk coexistence', $\mu = \mu_c$, as $b \to b_W^-$.

In order to analyze equation (\ref{kpzyv}) it is convenient to
perform a Cole-Hopf change of  variable $h({\bf x},t)=-\ln n({\bf x},t)$,
leading to
 \begin{equation}
\partial_t n=\nu \nabla^2 n-{\partial V(n)\over \partial n} + n \eta,
\label{toral}
\end{equation}
with $V(n)=an^2/2+ bn^3/3 +c n^4/4$.
This describes the interface problem as a diffusion-like equation with 
{\em multiplicative noise} \cite{MN,equiv}. 
In this
representation, the unbinding from the wall ($\langle h \rangle \to\infty$) 
corresponds to a transition into an {\it absorbing state} 
$\langle n \rangle \to 0$.  
In the following we will use both languages, $h$ and $n$, indistinctly
although the natural description of wetting is in terms of $h$.
The case $b>0$ was studied in \cite{MN,haye1}, while
$b<0$ is the case studied in \cite{nos,haye2,marsili,raul,muller}.

Note that we have made use of Ito calculus, and thus equation
(\ref{toral}) should be interpreted
in the Ito sense \cite{vankampen}. In general, potentials of the form
$b n^{p+2}/(p+2)+c n^{2p+2}/(2p+2)$ with $p>0$ result in 
equivalent effective Hamiltonians since, when expressed in 
terms of $h$, $p$ can be eliminated by redefining the height scale.
The case $p=2$ (with fixed $b<0$) has been studied in \cite{raul}
in the context of stochastic spatio-temporal intermittency (STI). The 
unsuspected connections between these two problems are illustrated in the
following sections.

\section{Mean field}
In this section we analyze equation (\ref{toral}) at the mean field level. 
We begin by discretizing (\ref{toral}) on a regular $d$-dimensional
lattice
\begin{equation}
\partial_t n_i= {\nu \over 2d} \sum_{j}(n_j-n_i)-
{\partial V(n_i) \over \partial n_i} + 
n_i \eta_i,
\label{disc}
\end{equation}
where $n_i=n(x_i,t)$ and the sum over 
$j$ runs over the nearest neighbors of $i$.
The Fokker-Planck equation for the one-site
stationary probability $P(n_i)$ can be easily worked-out.
In mean-field approximation (i.e. substituting the values
of the nearest neighbors by the average $n$ value), the
stationary Fokker-Planck equation for $D=1$ is 
\begin{equation}
{\partial\over \partial n}\bigg[
{\partial V(n)\over \partial n} +\nu (n-\n) P_t(n)\bigg]+
{\partial^2\over \partial n^2} \Big(n^2 P_t(n)\Big)=0
\label{fokkerplanck}
\end{equation}
and its associated solution
\begin{equation}
P(n,\n) = N {1 \over n^2} \exp -\int_0^n {V'(n)
+\nu (n-\n) \over n^2} dn,
\end{equation}
where the integration constant $N$ is determined by a normalization 
condition and $\n$ is obtained from the self-consistency requirement
\begin{equation}
\n={\int_0^\infty dn \ n P(n,\n) \over 
\int_0^\infty dn \ P(n,\n)} = F(\n).
\label{meanfield}
\end{equation}
Let us consider two limiting cases where analytic 
solutions of Eq.(\ref{meanfield}) can be worked out. 
In the zero-dimensional case, or equivalently $\nu=0$, the solution
of (\ref{fokkerplanck}) reads 
\begin{equation}
P(n)= N {\exp{-\Big( {b \over p}n^p+
{c \over 2p}n^{2p} \Big)} \over n^{a+2}},
\end{equation}
that, in terms of heights, is
$P(h) \sim \exp[(a+1)h - b  e^{-h}/p-c e^{-2h}/2p]$
and yields the effective potential $V_{eff}(h)=-\ln P(h)=
-(a+1)h + b  e^{-h}/p+c e^{-2h}/2p$. 
Clearly this coincides with the potential in (\ref{kpzyv}).
A {\it complete wetting} transition occurs when approaching $a=-1$ with
$b>0$ and critical wetting is found at $a=-1$ with $b=0$. 

For spatially extended systems, in the $\nu =\infty$ limit, a 
saddle-point expansion in $\nu$ yields 
$V'(n)=0$ \cite{vandenbroeck,marsili}. Thus, the dynamical behavior in
this limit is that of the deterministic mean-field version of
(\ref{fokkerplanck}): for any $p>0$, 
there is a line of second order wetting transitions 
at $a=0$ and $b>0$, and a line of first-order transitions at $a>0$ and 
$b=-(p+2) \sqrt{ac/(p+1)}$. These lines meet at a tricritical point 
at the origin.

\begin{figure}
\includegraphics[width=10cm]{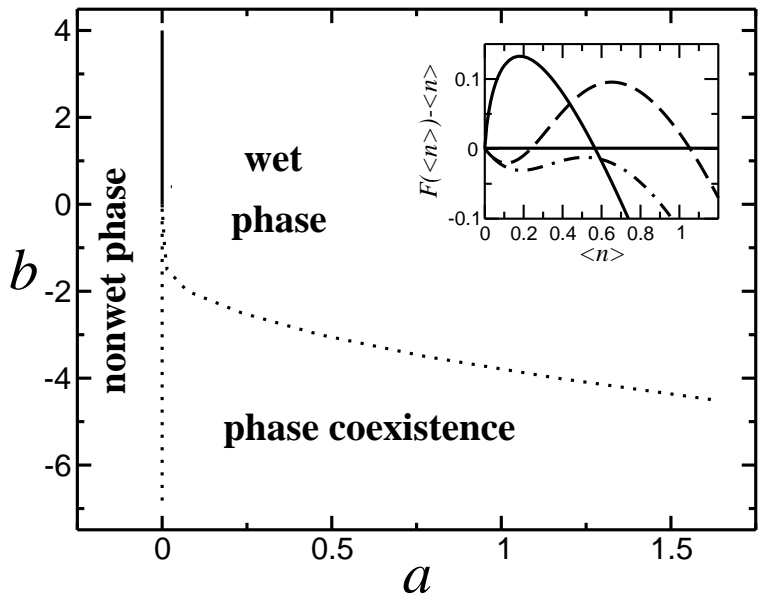}
\caption{Mean-field phase diagram and typical solutions of the equation 
$F(\n)=\n$ (temperature in units such that $k_B=1$.}
\label{pdandtipos}
\end{figure}

For values of $\nu$ other than zero or $\infty$, the
self-consistency equation $\n =F(\n)$ has to be solved numerically.
Without loss of generality, we set $p=2$, $c=1$,
and illustrate in Fig. \ref{pdandtipos} the three different regimes:  
one stable solution at $\n =0$ (dash-dotted line); one
unstable solution at
$\n =0$ and a stable one at $\n \not= 0$ (solid line); two stable 
solutions and an unstable one (dashed line). 
Stable solutions can be identified by a negative slope of 
$F(\n)-\n$ at the intersection point \cite{shiino}.  
A nonzero solution emerging continuously from $\n =0$ 
as a function of $a$ and $b$, signals a second order transition. 
This is the case for the dash-dotted and the solid lines in the 
inset of Fig. \ref{pdandtipos}.
By contrast, when the nontrivial solution appears discontinuously as a
function of $a$ and $b$, the transition is first-order
(dash-dotted and dashed lines). 
The corresponding phase diagram is depicted in Fig. \ref{pdandtipos}.
The solid line is a second-order phase boundary from non-wet to 
wet substrates. 
Between the two dotted lines, the wet and non-wet phases
coexist as stationary solutions of the dynamical equation. 
The three lines join at the tricritical point at $a=b=0$. 

In order to determine the order parameter critical 
exponent in mean field approximation, we proceed as in
\cite{genovese}. First, we rewrite (\ref{meanfield}) as
\begin{equation}
\n^{-1}=- \partial_{\n} 
\ln \int_0^\infty dt t^a \exp \bigg(-{b \over p} t^p
-{t^{2p} \over 2p}\bigg) e^{-\n t}.
\end{equation}
Next, we introduce a Gaussian transformation
%
and expand the resulting integrals for small $\n$. 
We find $ \n \sim |a|^{1/p}$, and thus $a_c=0$ and $\beta =1/p$. 
  
\section{Beyond mean-field theory}

In this section we explore whether the mean-field phase
diagram structure survives when the effects of fluctuations are taken
into account.
Mean-field exponents are expected to hold above the upper critical 
dimension $d_c$, which 
in the present case, equation (\ref{toral}), is known to be $d_c=2$,
(corresponding to 3 bulk dimensions and in the weak coupling
regime of the KPZ) \cite{MN}. 
For positive values of b and $d>2$, the second term $n^{2p+2}$ 
in the effective potential is irrelevant \cite{MN} 
and the we are left  with
\begin{equation}
\partial_t n= \nu \nabla^2 n -an -b n^{p+1}+ n\eta, 
\end{equation}
defining the {\em multiplicative-noise} (MN) universality 
class \cite{MN}. 

We have solved (\ref{toral})  numerically for 
different system dimensions.
In particular, for a one-dimensional 
substrate we have considered a system 
size $L=1000$, $\nu=p=2$, $D=1$ and $c=1.5$. 
The time step and the mesh size were set to 0.001 and 1, respectively. 
We started by determining the chemical potential
for which the free interface has zero average velocity.
For the parameters given above we found $a_c\approx -0.064$.
Then we fix $a=a_c$ and calculate $\langle n(t) \rangle$ for
different values of $b$ and large $t$. Length and time units are 
in lattice spacings and Monte Carlo steps, respectively.

\subsection{Critical wetting}

\begin{figure}
\includegraphics[width=8.5cm,angle=0]{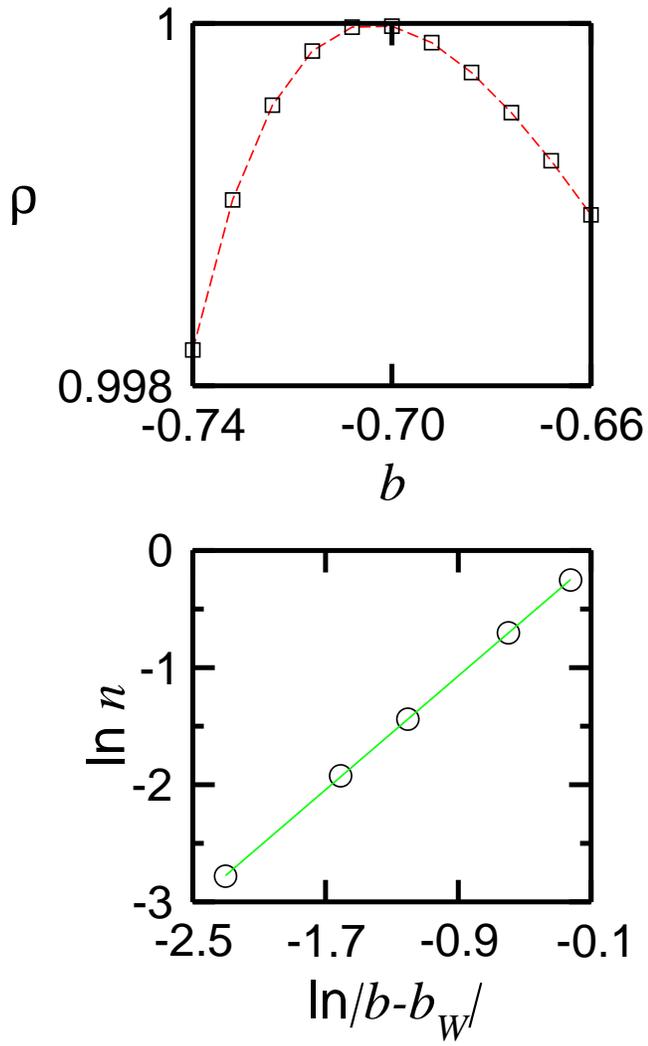}
\caption{Top: linear correlation coefficient for $\ln n$ as function of 
$\ln |b-b_W|$ for different 
values of $b_W$. The maximum gives the best estimate of $b_W=-0.70 \pm
0.01$. Bottom: log-log plot
of $\ln n$ as function of $\ln |b-b_W|$ in the vicinity of the critical point. 
The line is a least-squares fit; from its slope the critical wetting 
exponent is found to be $1.20 \pm 0.02$}
\label{critical}
\end{figure}

To study critical wetting we set $a=a_c$ and consider small values of
$b$ for which an initially pinned interface remains pinned, and increase
progressively $b$ until the non-wet phase becomes unstable at $b_W$. 
The critical point  
is estimated as the value $b_W$ that maximizes
the linear correlation coefficient of $\log \langle n \rangle$ 
versus $\log|b-b_W|$; the critical exponent is then determined from 
the corresponding slope (Fig. \ref{critical}).
It is found that the critical ``temperature'' is
depressed from its mean-field value $b_W=0$ to $b_W= -0.70 \pm 0.01$,
with an  associated critical exponent $\beta =1.20 \pm 0.01$
(the error in the exponent comes from a least-squares fit).
Below (above) that value we find first (second) order depinning transitions,
by varying $a$.
Therefore, as in mean field, there is a ``tricritical'' point, 
joining a line of second  order transitions ($b>b_W$) 
with one of first order transitions ($b<b_W$). 
{\it The critical exponents and universality of this 
multiplicative-noise tricritical point has not been investigated before}. 

The finite coexistence region allows us to define
critical wetting along a range of different paths, delimited by the 
dashed lines in the mean-field diagram 
of Fig. \ref{pdandtipos}. We have checked that the
value of $\beta$ does not change when the critical 
point is approached along different paths within this region.
Further numerical and analytical studies of this new 
universality class will be left for future work.

\subsection{Complete Wetting,  $b >b_W$ case}

We consider a one-dimensional substrate, with $b=1>b_W$, let the system
evolve to the stationary state and 
then compute the order parameter $\langle n \rangle$ 
for different values of $a$ near its critical value. 
As $a \to a_c$ a continuous transition into
an absorbing state $n=0$ is observed; it is the non-equilibrium
counterpart of complete-wetting.
The associated critical exponent is found to  be $\beta =1.65 \pm 0.05$ 
in good agreement with the prediction for
the MN class, $\beta=1.5 \pm 0.1$ \cite{genovese}.  
Other positive values of $b$ yield similar results.
In addition, we have simulated systems above the upper 
critical dimension, in $d=3$, 
with $L=25$, $b=5$ and other parameters as 
in the one-dimensional case. 
Our best estimate for $\beta$ is $\beta=0.96 \pm 0.05$,
indicating that this transition 
is governed by the weak noise fixed point of the
MN class \cite{MN}. For larger values of the noise
amplitude we find a strong coupling  transition, in agreement
with the theoretical predictions \cite{MN,genovese}. 
Finally, we note that both numerical and renormalization group arguments 
lead to $\beta=1$ in the weak coupling  regime,
independently of the value of $p$ \cite{MN}. 
This is at odds with the 
mean-field prediction $\beta=1/p$. 
Birner et al. \cite{birner} have recently suggested a transition 
from $1/p$ to a nonuniversal behavior depending on the ratio of the
noise to the strength of the spatial coupling. However, this discrepancy 
appears to be generic since different types of mean-field approaches 
yield the same (incorrect) result and the origin of the discrepancy
remains unclear \cite{genovese}.

Since the Cole-Hopf transform of the MN equation is the
same as KPZ with an additional exponential term, the
MN exponents are those of KPZ iff the extra term
is an irrelevant term of the KPZ renormalization group flow.
Note that the Cole-Hopf transform fixes the value of 
$\lambda /\nu$ and thus the $\lambda =0$ limit can not be considered when 
this transformation is used.
In addition the potential $V(h)$ is a relevant term of the EW equation.
In this regime adding a non-linear potential is a relevant perturbation
and it does indeed change/determine the wetting exponents (cf. with the
literature on equilibium wetting \cite{margarida}).
Thus {\em EW plus a (non-linear) wetting potential is not
equivalent to KPZ in the weak coupling regime plus the same wetting
potential}.

\subsection{Depinning transition at $b <b_W$}

\begin{figure}
\includegraphics[width=15cm]{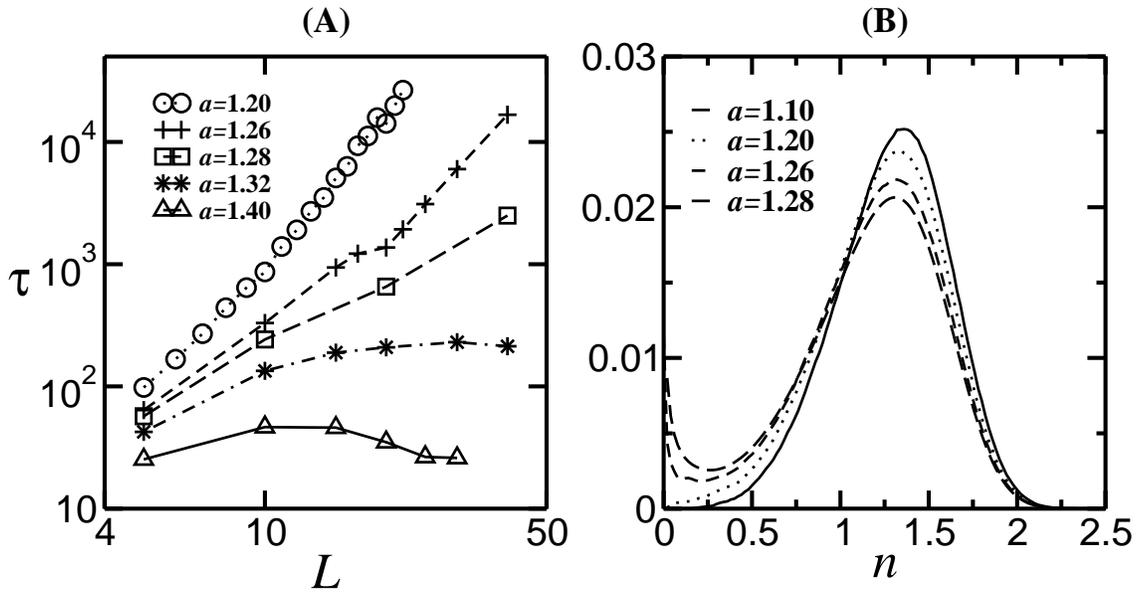}
\caption{(A) Characteristic depinning times 
and (B) ss-pdf for various representative values of $a$.}
\label{histo}
\end{figure}

As expected, no transition is found as $a
\to a_c$ when equation (\ref{toral}) is solved numerically
for $b<b_W$. Of course, the system undergoes a 
pinning/depinning transition when crossing the $a_c=0$
boundary line, but this transition is driven by the chemical
potential difference rather than by the substrate potential 
and thus it is unrelated to wetting, where phase
coexistence of the ``liquid'' and ``gas'' phases is
required (i.e. $a=a_c$). A very 
rich phenomenology associated with these transitions.
have been found, however.
For $b=-4$ we find that the non-wet phase becomes unstable
at $a^* \approx 1.3$ and that the wet phase becomes unstable
at $a_c \approx -0.064$ (as before). 
Consequently, in the range $a_c < a < a^\ast$ 
both phases coexist. This means that if the interface is initially close
to the wall ($n > 1$) it remains pinned, while if it is initially far from 
the wall ($n \lappeq 1$) it detaches and moves away with a constant velocity. 
Therefore, 
the system undergoes a first-order transition as a function of
$a$. 
In order to establish the phase boundaries we have used the following  
criteria.
The stability of the pinned phase may be characterized by the
time $\tau$ taken by the interface to depin in the limit 
$L \to \infty$. $\tau$ can be defined as the time taken by 
the last site of the interface to detach, $h({\bf x})>0$ or 
$n({\bf x})<1$ $\forall {\bf x}$. Similarly, we may define $\tau$ 
as the time characterizing the asymptotic exponential decay of 
$\langle n(t) \rangle$, 
where the angular brackets denote averages of
independent runs ( typically $10^5-10^6$ in our simulations). 
We have verified that both definitions yield analogous results.
As shown in Fig. \ref{histo}A, for $a>a^*$, $\tau$ saturates with 
increasing system size and thus the interface detaches in a finite time.
Within the coexistence region we have found two different
regimes: close to the stability threshold of the pinned phase 
there is a narrow
stripe $1.22 \lappeq a \lappeq 1.3$ where the detaching time 
grows approximately as a power-law. For 
$a_c \lappeq a \lappeq 1.22$ $\tau$ grows exponentially with $L$. 
In both cases $\tau$ diverges as $L \to \infty$, implying that
the pinned phase is stable in the thermodynamic limit. 
Due to the very large characteristic times, we cannot discard the
possibility that the power-laws are also (asymptotically) exponentials.
The study of the asymptotic behavior of the detaching times
requires longer simulations, beyond our current computer capabilities. 
Finally, the non-monotonic behavior of the characteristic times, 
as well as the step in the curve for $a=1.28$, may be accounted for by 
the presence of two different competing mechanisms as  
described in \cite{nos}: once a site is detached it pulls out its
neighbors which, in turn, pull out their neighbors in a cascade effect
until the whole interface is depinned in a time which grows linearly 
with $L$. This is more likely for small systems, but the probability
that a site gets detached increases with the system size. 

Another way to characterize the power-law regime
is to analyze the single-site stationary probability 
density function (ss-pdf), defined as the average of 
$n(t)$ over pinned states rather than over all runs. 
Figure \ref{histo} shows the unnormalized ss-pdf 
for different values of $a$. In the exponential-regime 
($a<1.22$) the histogram exhibits a maximum at a pinned state with
$\langle n \rangle > 0$.
In the power-law-regime, however, the histogram develops
a secondary maximum near $n=0$, indicating that a  
fraction of the interface depins. As $a$ increases,
the secondary maximum, at zero $n$, increases 
while the maximum, at finite $n$, decreases. At the stability 
edge ($a^*\approx 1.3$) the histogram changes abruptly into a delta 
function at $n=0$ and the pinned phase becomes unstable.

\begin{figure}
\includegraphics[width=12cm]{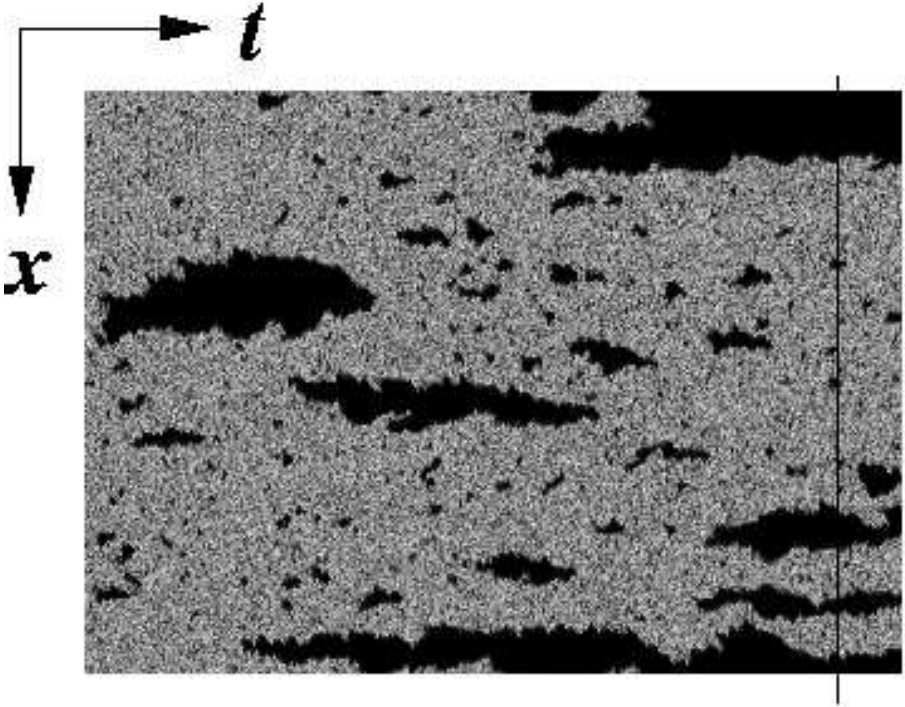}
\caption{Configuration in the $n$-representation for $a=1.18$ and
$b=-4$. Depinned regions
($n<1$) are colored in dark-grey and pinned ones ($n>1$) in light-gray. 
1000 time slices are depicted at intervals of 50 time uints. 
The system size is $L=500$.}
\label{confn}
\end{figure}

The differences between the exponential and power-law regimes
are also observed in a space-time snapshot of 
a numerical solution of (\ref{toral}). In Fig. \ref{confn}
we plot the stationary field $n$, for $a=1.28$, exhibiting patterns
characteristic of STI \cite{raul}. 
The main feature of these patterns is
the appearance of depinned patches (absent for values of $a$ 
in the exponential regime) with a 
wide range of sizes and life-times
within the pinned phase. This regime, overlooked 
in previous studies of nonequilibrium depinning 
transitions \cite{haye2,marsili}, seems  
to correspond to the power-law regime described earlier. It is therefore
restricted to a narrow range between the exponential 
and the depinned regimes. 
This finding is at odds with results of previous
work claiming that STI is generic in the 
coexistence region \cite{raul}.     

\begin{figure}
\includegraphics[width=8cm,angle=270]{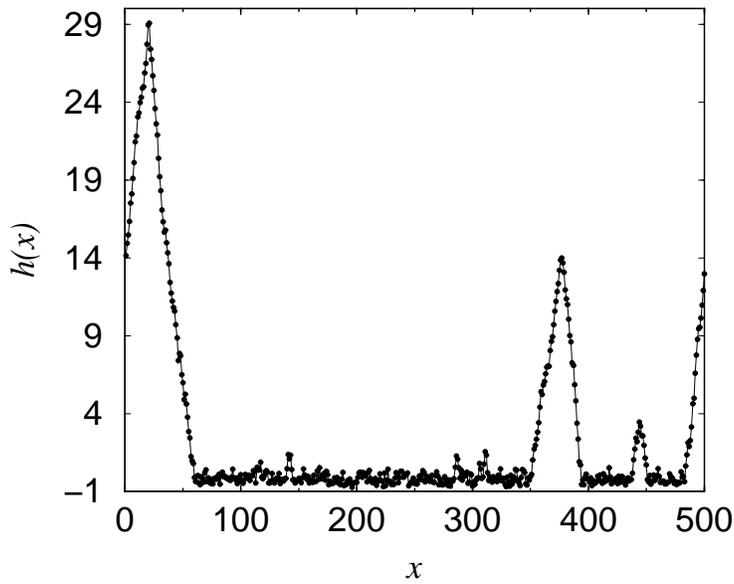}
\caption{Instantaneous configuration of the interface for time slice 400 
(marked with a line in 
Fig. \ref{confn}). Parameters as in Fig. \ref{confn}.}
\label{confh}
\end{figure}

A typical profile in terms of $h$ is shown in Fig. \ref{confh}. 
The depinned interfacial regions form triangles   
with constant average slope $s$. These triangular droplets 
are similar to those described in the discrete model 
of \cite{haye2}. 
By taking averages of (\ref{kpzyv}),  the typical slope, $s$,
 of the triangular facets is determined through 
\begin{equation}
|\lambda_R| s^2 =a+1,
\label{slopes}
\end{equation}
where $\lambda_R$ is the renormalized non-linear 
coefficient of the KPZ equation. 
In order to verify equation (\ref{slopes})
we have fixed $\nu=p=D=1$, $\lambda=-1$
and $a=2.16$. Averaging over 250.000 different triangles, we
find an average slope $s=1.781$, 
while the value of $\lambda_R$ calculated from 
the tilt-dependent velocity of the depinned interface \cite{barabasi}
 yields $\lambda_R=-0.9934$ from which $s=1.784$, 
in excellent agreement with the previously measured value.
  
\begin{figure}
\includegraphics[width=15cm]{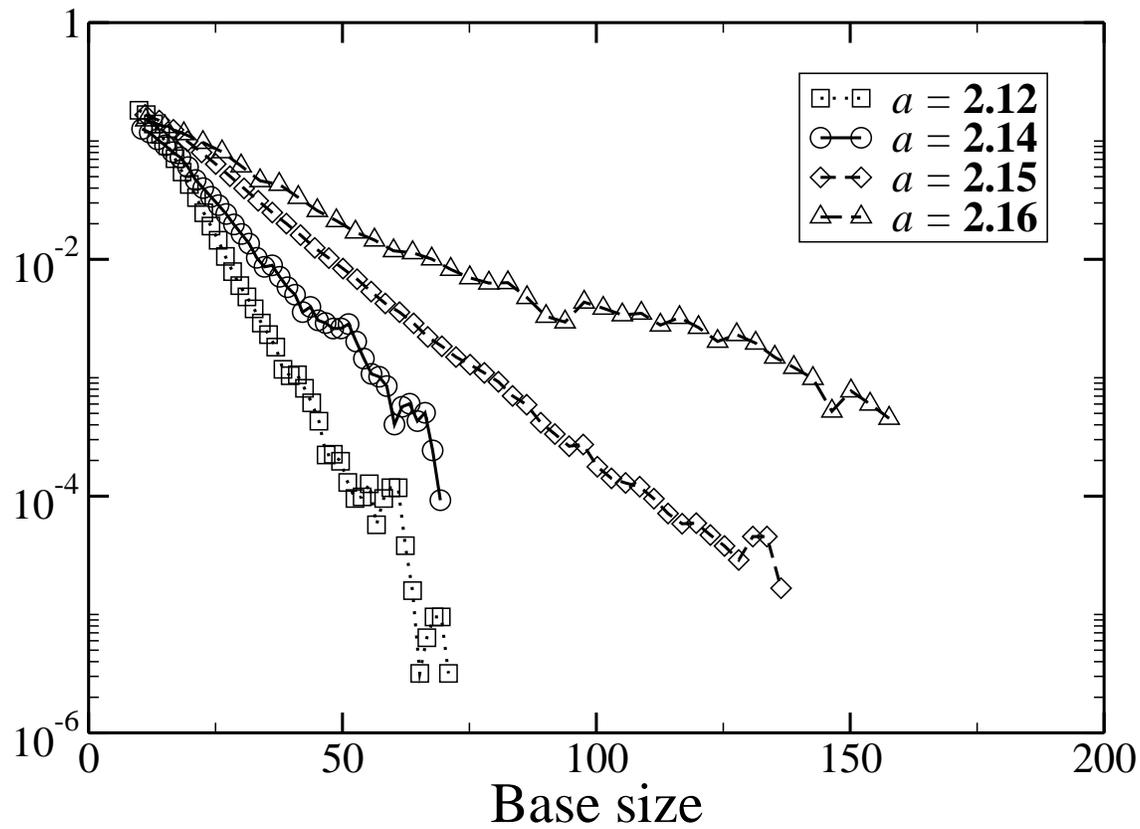}
\caption{Distribution of triangles as a function of the size of the
triangular base, for
$a=2.12,2.14,2.15$ and 2.16.}
\label{tridistri}
\end{figure}

We  also studied the size distribution of triangles within
the power-law regime. Our results correspond to 
$\nu=p=D=1$, $\lambda=-1$, $L=500$, $b=-4$, and the following
values of $a$: $2.12, 2.14, 2.15$, and $2.16$, 
and are summarized in Fig. \ref{tridistri}.
$a \approx 2.10$ is the boundary between 
the power-law and exponential regimes 
and the pinned phase is unstable for $a^* \approx 2.18$.
The maximum size of the depinned regions increases as this
instability is approached. Our
data suggests an exponential dependence on the size of the 
triangular base.
This indicates that there is a maximum size for the depinned regions
and thus rigorous scale invariance (typical of growth
driven by a coarsening mechanism) of the STI region is ruled out.
More explicitly, the distribution of
triangle sizes, $l$, is described very well by the function
$\exp [3.44(a-2.176)l]$, implying that the exponential slopes 
in Fig. \ref{tridistri} are proportional to $a-a^*$.
Clearly, triangles with a base less than $\sim 2, 3$ cannot be
visualized due to the discretization of equation (\ref{toral}).
A simple extrapolation indicates that the triangles become 
imperceptibly small for $a \approx 2.05$, in good agreement with the
value obtained for the boundary between 
the power-law and exponential regimes. 
Therefore, we cannot rule out the possibility 
that the triangles are ubiquitous throughout the coexistence region
(although not always visible in a discrete numerical simulation)
 in
which case the power-law dettaching times should turn into exponentials
for large enough times and system sizes.  In this case the force exerted
by the non-linear KPZ term on the triangular facets against the 
direction of growth, at late times, guarantees the stability 
of the pinned phase \cite{haye2} throughout the finite coexistence region. 

\begin{figure}
\includegraphics[width=9cm]{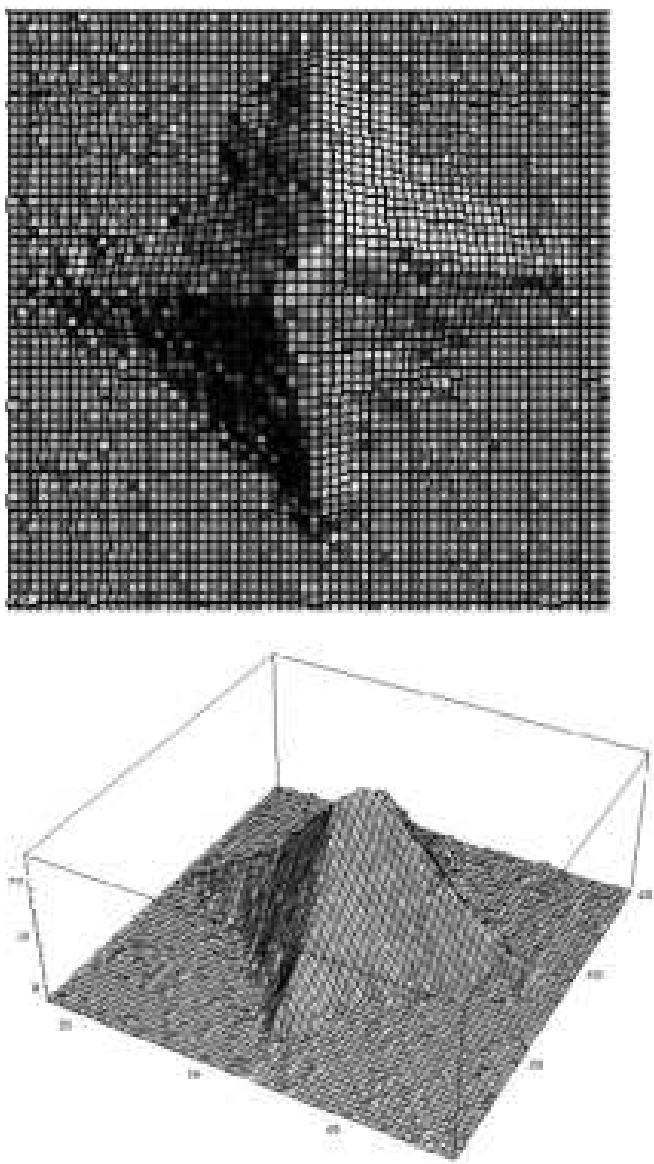}
\caption{Snapshot of an interface configuration for a 
$100 \times 100$ system (not all the substrate is shown) 
and parameters $a=2$ and $b=-4$. } 
\label{pyramid}
\end{figure}

Finally, we study the phase-coexistence regime in
a two-dimensional system to check whether the triangular
patterns survive in higher dimensionalities. 
In particular, we consider a system 
size $100 \times 100$ and take $a=2$, $b=-4$, well within the coexistence
region. 
We find structures as those shown in Fig. \ref{pyramid}: 
the triangles becoming pyramids. Note that the edges of the pyramid
bases are parallel to the axes of the discretization-lattice.
This suggests that the pyramids are lattice artefacts and that a
continuum system may exhibit conical structures. 

\section{Conclusions}

We have investigated a continuum model for nonequilibrium wetting
transitions. The model consists of a KPZ equation in the presence
of  a short-ranged substrate potential, 
and is the most natural non-equilibrium extension of the interface 
displacement models used in equilibrium wetting.
It can be mapped into a multiplicative noise problem, 
enabling simple theoretical calculations at the mean-field level. 
Numerical simulations reproduce a phase diagram analogous to that
obtained within mean-field, including first as well as second-order
phase transitions.
In particular, we have found complete wetting and critical wetting
transitions, as well as a finite area in the temperature-chemical
potential phase diagram where pinned and depinned phases coexist.
This finite coexistence region allows us to define
critical wetting along a range of paths that are, however, characterized
by the same critical exponents.  
Within this area we identified two regimes. In the first, the lifetime of 
the pinned phase grows exponentially with increasing system size and 
its ss-pdf is bell-shaped.  
The second one exhibits STI, lifetimes consistent with a 
power-law, and a double-peaked ss-pdf. The main feature of the 
latter regime is the presence of triangular structures that
have been characterized by their slopes and size distributions. 

An interesting  open problem is that of the equilibrium limit 
of non-equilibrium wetting. 
The Cole-Hopf transform precludes the limit 
$\lambda =0$ to be studied using this method. 
Moreover, we have noted how
the behavior of the EW equation in the presence of a wetting potential
differs from the weak-noise regime of the MN equation. 
This leaves the crossover to equilibrium wetting an open 
challenge. In addition, the effects of long-ranged potentials 
on the phenomenology described here remain to be investigated.

Finally it would be extremely interesting to develop experiments 
in order to explore the rich, non-equilibrium phenomenology described in
the previous sections; 
liquid-crystals are in our opinion good candidates for this. 
It is our hope that this work will stimulate experimental studies
in this direction.

\section{acknowledgments}  
We acknowledge financial support from the E.U. through Contracts No. 
ERBFMRXCT980171 and ERBFMRXCT980183, by the Ministerio de Ciencia y
Tecnolog\'\i a (FEDER) under project BFM2001-2841 and from the 
Funda\c c\~ao para a Ci\^encia e a Tecnologia, contract SFRH/BPD/5654/2001.

\newpage

\newpage

\newpage

\newpage

\newpage

\newpage

\begin{thebibliography}{99}
\bibitem{nota2}If, instead, one considers long-range (van der Waals) interactions,
the potential has the general form \cite{margarida}.
\begin{equation}
V(h)=b(T) h^{-m}+c h^{-n}, \qquad n>m>0.
\end{equation}
\bibitem{note}
The influence of the sign of $\lambda$ 
has been previously investigated in \cite{haye1,haye2,MN}.
\end{thebibliography}
\end{document}